\documentclass{article}
\usepackage[T1]{fontenc}
\usepackage{graphics}

\makeatletter

\newcommand{\LyX}{L\kern-.1667em\lower.25em\hbox{Y}\kern-.125emX\@}

\usepackage{a4}
\input{epsfig.sty}
\def\fnum@table{\tablename~{\bf\thetable}}
\def\fnum@figure{\figurename~{\bf\thefigure}}
\def\tablename{\footnotesize{\bf Table}}
\def\figurename{\footnotesize{\bf Figure}}

\def\be{\begin{equation}}
\def\ee{\end{equation}}

\makeatother

\begin{document}

\title{\textbf{Cosmic Ray Air Shower Characteristics }\\
\textbf{in the Framework  of the }\\
\textbf{Parton-Based Gribov-Regge Model} \textbf{\textsc{\Large NE}}\textbf{\textsc{X}}\textbf{\textsc{\Large US}}\textbf{\textsc{}}\\
\textbf{\textsc{}}\\
}

\author{\textbf{G. Bossard\protect\( ^{1}\protect \), H.J. Drescher\protect\( ^{1}\protect \),
N.N. Kalmykov\protect\( ^{2}\protect \), S. Ostapchenko\protect\( ^{2,1}\protect \),
}\\
\textbf{A.I. Pavlov\protect\( ^{2}\protect \),T. Pierog\protect\( ^{1}\protect \),
E.A. Vishnevskaya\protect\( ^{2}\protect \),  K. Werner}\protect\( ^{1}\protect \)\\
\\
\\
 \textit{\protect\( ^{1}\protect \)} \textit{\small SUBATECH, Université de
Nantes -- IN2P3/CNRS -- Ecole des Mines,  Nantes, France }\\
\textit{\small \protect\( ^{2}\protect \) Skobeltsyn Institute of Nuclear Physics,
Moscow State University, Moscow, Russia}\textit{}\\
}

\maketitle
{\par\centering \textbf{\large Abstract}\large \par}

The purpose of this paper is twofold: first we want to introduce a new type
of hadronic interaction model (\textsc{neXus}), which has a much more solid
theoretical basis as, for example, presently used models like \textsc{qgsjet}
and \textsc{venus}, and ensures therefore a much more reliable extrapolation
towards high energies. Secondly, we want to promote an extensive air shower
(EAS) calculation scheme, based on cascade equations rather than explicit Monte
Carlo simulations, which is very accurate in calculations of main EAS characteristics
and extremely fast concerning computing time. We employ the \textsc{neXus} model
to provide the necessary data on particle production in hadron-air collisions
and present the average EAS characteristics for energies \( 10^{14}-10^{17} \)
eV. The experimental data of the \textsc{casa-blanka} group are analyzed in
the framework of the new \textsc{}model.

\section{Introduction\label{intro.sec} }

Although cosmic rays have been studied for many decades, there remain still
many open questions, in particular concerning the high energy cosmic rays above
\( 10^{14} \) eV. One does neither know their composition nor the sources and
acceleration mechanisms, partly because of the fact that at these high energies
direct measurements are impossible due to the weak flux. But since cosmic ray
particles initiate cascades of secondaries in the atmosphere, the so-called
extensive air showers, one may reconstruct cosmic ray primaries by measuring
shower characteristics. This reconstruction requires, however, reliable model
predictions for the simulation of EAS initiated by either protons or nuclei
from helium to iron. The problem is that the energy of the cosmic rays may exceed
by far the energy range accessible by modern colliders, where at most equivalent
fixed target energies of roughly \( 10^{15} \) eV can be reached. The projects
of new generation EAS arrays are aimed even to the energy region \( 10^{20}\div 10^{21} \)~eV,
and so the gap between the existing energy limit of collider data and the demands
of cosmic ray experiments is considerable. Moreover, the real reliable data
limits are essentially less than above mentioned, because at present time collider
experiments do not register particles going into the extreme forward direction
and a number of other drawbacks may be listed. Therefore there exists a real
need of ``reasonable'' models, implementing the correct physics, in order
to be able to make extrapolations towards extremely high energies. 

Concerning models one has to distinguish between EAS models and hadronic interaction
models. The latter ones like \textsc{venus} \cite{wer01}, \textsc{qgsjet} \cite{qgsa02,qgs02},
and \textsc{sybill} \cite{syb03} are modeling hadron-hadron, hadron-nucleus
and nucleus-nucleus collisions at high energies, being more or less sophisticated
concerning the theoretical input, and relying in any case strongly on data from
accelerator experiments. The EAS models like \textsc{corsika} \cite{cor04}
are actually simulating the full cascade of secondaries, using one of the above-mentioned
hadronic models for the hadronic interactions and treating the well known electro-magnetic
part of the shower. It turns out that the model predictions of EAS simulations
depend substantially on the choice of the hadronic interaction model. In \textsc{corsika,}
the average electron number in EAS at primary energy \( 10^{15} \)~eV varies
from \( 1.11\cdot 10^{5} \) to \( 1.62\cdot 10^{5} \) (at sea level) depending
on the hadronic interaction model \cite{heck05}. So the right choice of the
model and its parameters is extremely important. 

Recently a new hadronic interaction model \textsc{neXus} \cite{dre00} has been
proposed. It is characterized by a consistent treatment for calculating cross
sections and particle production, considering energy conservation strictly in
both cases (which is not the case in all above-mentioned models!). In addition,
one introduces hard processes in a natural way, avoiding any unphysical dependence
on artificial cutoff parameters. A single set of parameters is sufficient to
fit many basic spectra in proton-proton and lepton-nucleon scattering, as well
as in electron-positron annihilation. Briefly: concerning theoretical consistency,
\textsc{neXus} is considerably superior to the presently used approaches, and
allows a much safer extrapolation to very high energies. 

This new approach cures some of the main deficiencies of two of the standard
procedures currently used: the Gribov-Regge theory and the eikonalized parton
model, the theoretical basis of the above-mentioned interaction models. There,
cross section calculations and particle production cannot be treated in a consistent
way using a common formalism. In particular, energy conservation is taken care
of in case of particle production, but not concerning cross section calculations.
In addition, hard contributions depend crucially on some cutoff and diverge
for the cutoff being zero.

Having a reliable hadronic interaction model, one may now proceed to do air
shower calculations. It may appear that an ideal solution from the user's point
of view is to use direct Monte Carlo technique, where the cascade is traced
from the initial energy to the threshold one and the threshold energy corresponds
to the minimum energy registered by the array in question. But such an approach
takes unreasonably much computer time and sometimes gives no possibility to
analyze experimental data in the appropriate way. Already at \( 10^{14} \)~eV
to \( 3\cdot 10^{16} \)~eV, the experimental statistics from the \textsc{kascade}
experiment \cite{kas07} is ten times bigger than the corresponding number of
simulated events. It should be mentioned that at higher energies (above \( 10^{17} \)
eV) it is hardly possible to use direct Monte Carlo calculations. Of course,
some modifications, like the thinning method~\cite{hil08}, are possible and
average values may be computed.

In this paper, we follow an alternative approach, where one treats the air shower
development in terms of cascade equations. Here, the cascade evolution is characterized
by differential energy spectra of hadrons, which one obtains by solving a system
of integro-differential equations. Crucial input for these equations is the
inclusive spectra of hadrons produced in hadron-air collisions. These spectra
are obtained by performing \textsc{neXus} simulations and finally parameterizing
the results. Having solved the cascade equations, we finally analyse the results
of the \textsc{casa-blanka} group as a first application of our approach.

\section{The NE{\LARGE X}US Model \label{NM label}}

The most sophisticated approach to high energy hadronic interactions is the
so-called \textbf{Gribov-Regge theory} \cite{gria09,grib09}. This is an effective
field theory, which allows multiple interactions to happen ``in parallel'',
with phenomenological object called ``Pomerons'' representing elementary interactions
\cite{bak76}. Using the general rules of field theory, one may express cross
sections in terms of a couple of parameters characterizing the Pomeron. Interference
terms are crucial, they assure the unitarity of the theory. 

A big disadvantage is the fact that cross sections and particle production are
not calculated consistently: the fact that energy needs to be shared between
many Pomerons in case of multiple scattering is well taken into account when
considering particle production (in particular in Monte Carlo applications),
but not for cross sections \cite{abr92}. 

Another problem is the fact that at high energies, one also needs a consistent
approach to include both soft and hard processes. The latter ones are usually
treated in the framework of the parton model, which only allows to calculate
inclusive cross sections.

\begin{figure}[htb]
{\par\centering \resizebox*{!}{0.25\textheight}{\includegraphics{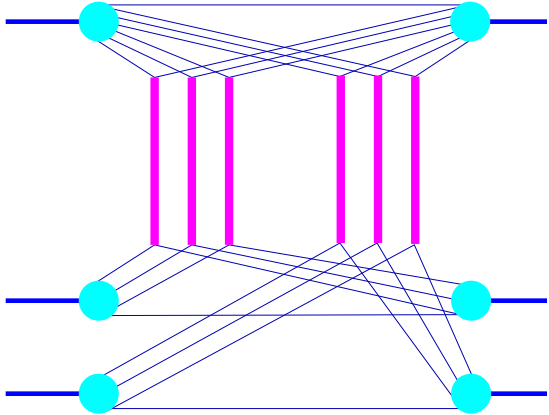}} \par}

\caption{The diagram representing a proton-nucleus collision, or more precisely a proton
interacting with (for simplicity) two target nucleons, taking into account energy
conservation. Here, the energy of the incoming proton is shared between all
the constituents, which provide the energy for interacting with two target nucleons.\label{grtppa}}
\end{figure}

We recently presented a completely new approach \cite{nex06,dre99a,dre00} for
hadronic interactions and the initial stage of nuclear collisions, which is
able to solve several of the above-mentioned problems. We provide a rigorous
treatment of the multiple scattering aspect, such that questions of energy conservation
are clearly determined by the rules of field theory, both for cross section
and particle production calculations. In both (!) cases, energy is properly
shared between the different interactions happening in parallel, see fig. \ref{grtppa}
for hadron-nucleus collisions. This is the most important and new aspect of
our approach, which we consider a first necessary step to construct a consistent
model for high energy nuclear scattering. 

The elementary diagram, shown as the thick lines in the above diagrams, is the
sum of the usual soft Pomeron and the so-called semi-hard Pomeron, where the
latter one may be obtained from perturbative QCD calculations (parton ladders).
To some extent, our approach provides a link between the Gribov-Regge approach
and the parton model, we call it ``Parton-based Gribov-Regge Theory''.

\section{System of Hadronic Cascade Equations\label{shce.label}}

EAS are produced as a result of the hadronic cascade development in the atmosphere.
We characterize hadronic cascades by the differential spectra \( h_{n}(E,X) \)
of hadrons of type \( n \) with energy \( E \) at an atmospheric depth \( X \),
the latter one being the integral over the atmospheric density \( \rho  \)
along a straight line trajectory (not necessary radial) from some point \( P \)
to infinity, 
\begin{equation}
X=\int _{P}^{\infty }\, \rho (x)dx,
\end{equation}
 measured usually in \( \mathrm{g}/\mathrm{cm}^{2} \). The decrease of the
average hadron numbers due to collisions with air nuclei is given as
\begin{equation}
\frac{dh_{n}}{dX}=-\frac{h_{n}}{\lambda _{n}},
\end{equation}
 where the mean inelastic free path \( \lambda _{n} \) (in units of mass/area)
can be expressed via the average hadron-air cross-section \( \sigma _{\mathrm{inel}}^{(n)} \)
and the average mass of air molecules \( m_{air} \) :
\begin{equation}
\lambda _{n}=\frac{m_{\mathrm{air}}}{\sigma ^{(n)}_{\mathrm{inel}}}
\end{equation}
 The second process to be considered is particle decay. The decay rate in the
particle c.m. system is \( dh_{n}/d\tau =-h_{n}/\tau _{0} \), with \( \tau _{0} \)
being the particle life time. For a relativistic particle we find
\begin{equation}
\frac{dh_{n}}{dX}=-\frac{B_{n}}{EX}h_{n},
\end{equation}
with the decay constant in energy units
\begin{equation}
B_{n}=\frac{m_{n}}{\alpha c\tau _{0}},
\end{equation}
where \( m_{n} \) is the hadron mass and \( c \) the velocity of light. The
ratio \( \alpha =\rho /X \) depends only weakly on \( X \) and is for the
following taken to be constant, which implies constant \( B_{n} \). If we take
the simple exponential barometric formula for the density \( \rho  \), we get
\begin{equation}
\alpha =\frac{\rho _{0}\, g\, \cos \theta }{P_{0}},
\end{equation}
where \( \rho _{0} \) and \( P_{0} \) are density and pressure at (for example)
sea level, \( g \) the gravitational acceleration of the earth, and \( \theta  \)
the zenith angle of the shower trajectory. In this paper, we only consider the
case \( \theta =0 \). 

Based on the above discussion, a system of integro-differential equations for
the differential energy spectra \( h_{n} \) of hadrons may be presented as
\begin{eqnarray}
\frac{\partial h_{n}(E,X)}{\partial X} & = & -h_{n}(E,X)\left[ \frac{1}{\lambda _{n}(E)}+\frac{B_{n}}{EX}\right] \label{sys.1} \\
 &  & +\sum _{m}\int _{E}^{E_{\mathrm{max}}}h_{m}(E',X)\left[ \frac{W_{mn}(E',E)}{\lambda _{m}(E')}+\frac{B_{m}D_{mn}(E',E)}{E'X}\right] dE',\nonumber 
\end{eqnarray}
 The quantities \( W_{mn}(E',E) \) and \( D_{mn}(E',E) \) are the inclusive
spectra of secondaries of type \( n \) and energy \( E \) which are produced
in interactions (\( W \)) or decays (\( D \)) of primaries of type \( m \)
and energy \( E' \). The energy \( E_{\mathrm{max}} \) is the maximum energy
considered. If the type of primary hadron is \( n_{o} \), its energy \( E_{0} \)
and the cascade originates at \( X_{o} \), then one should add 
\begin{equation}
h_{n}(E,X=X_{o})=\delta _{nn_{o}}\delta (E-E_{0})
\end{equation}
as the boundary condition. The more detailed consideration of the problem may
be found elsewhere (see \cite{gai10}). 

It is reasonable to incorporate in the system nucleons (and anti-nucleons),
charged pions (\( B_{\pi }=114 \)~GeV), charged kaons (\( B_{K}=852 \)~GeV),
and neutral kaons (\( B_{K^{o}_{L}}=205 \)~GeV). As \( B_{K^{o}_{S}}=1.19\cdot 10^{5} \)~GeV,
there is no sense to account for neutral kaons \( K_{o}^{S} \) at energies
\( \ll B_{K^{o}_{S}} \). But these particles should be included if their energy
exceeds \( 0.01B_{K^{o}_{S}} \). The values of decay constants are given at
the height \( 11 \)~km.

The computational technique to solve the system (\ref{sys.1}) is based on the
same principle as the traditional approaches \cite{ded11, hil12}, but some
improvements are introduced, which enable one to avoid too small steps when
integrating over the depth \cite{KaM13}. One discretizes the energy as
\begin{equation}
\label{sys.7}
E_{i}=E_{\mathrm{min}}\cdot c^{i}
\end{equation}
 with \( E_{\mathrm{min}}\simeq 1 \)~GeV and \( c \) such that the number
of points per order of magnitude is \( 10\div 20 \). Replacing the integral
in the right-hand side of (\ref{sys.1}) by the corresponding sum, one may write
\begin{equation}
\label{sys.5}
\frac{\partial h_{ni}(X)}{\partial X}=-h_{ni}(X)\left[ \frac{1}{\lambda _{ni}}+\frac{B_{n}}{E_{i}X}\right] +\sum _{m}\sum _{j_{min}}^{j_{\mathrm{max}}}h_{mj}(X)\left[ \frac{W_{mn}^{ji}}{\lambda _{mj}}+\frac{B_{m}D_{mn}^{ji}}{E_{j}X}\right] \, ,
\end{equation}
 with \( h_{ni}(X)= \)\( h_{n}(E_{i},X) \), \( \lambda _{ni}=\lambda _{n}(E_{i}) \),
and
\begin{eqnarray}
W_{mn}^{ji} & = & \int _{E_{\mathrm{min}}(i)}^{E_{\mathrm{max}}(i)}\frac{E}{E_{i}}W_{mn}(E_{j},E)dE\\
D_{mn}^{ji} & = & \int _{E_{\mathrm{min}}(i)}^{E_{\mathrm{max}}(i)}\frac{E}{E_{i}}D_{mn}(E_{j},E)dE
\end{eqnarray}
with \( E_{\mathrm{min}}(i)=E_{i}/\sqrt{c} \) and \( E_{\mathrm{max}}(i)=E_{i}\cdot \sqrt{c} \)
( if \( i=j \) then \( E_{max}(i)=E_{j} \) ). The factor \( E/E_{i} \) has
been added in the integral to ensure exact energy conservation. The spectra
\( W_{mn}(E_{j},E) \) and \( D_{mn}(E_{j},E) \) must be provided in order
to calculate the above integrals. The standard recipe for calculating \( D_{mn}(E_{j},E) \)
may be found elsewhere ( see \cite{gai10} or \cite{cor04} ) and implies no
difficulties. The calculation of \( W_{mn}(E_{j},E) \) must be based on \textsc{neXus}
simulations, as discussed below in detail. Since the \( E_{j} \) dependence
is very smooth, there is no point to calculate the spectra for all energies
\( E_{j} \). One rather chooses some reference energies \( E_{j'} \) to calculate
the spectra based on \textsc{neXus} simulations, and then interpolates to obtain
the spectra for the other energies. This method has proven to be superior concerning
the computational time when compared with the direct spectra calculations over
all energies. For the calculations in this paper we choose two reference points
per order of magnitude, starting from \( 10^{11} \)\( \,  \)eV:  \( E_{j'}=10^{11} \)\( \,  \)eV,
\( 10^{11.5} \)\( \,  \)eV, \( 10^{12} \)\( \,  \)eV etc. up to \( 10^{17} \)\( \,  \)eV
which is at present the maximum energy attainable in the \textsc{neXus} model.
As the \textsc{neXus} model is not valid at energies below \( 10^{11} \)\( \,  \)eV,
the data obtained with other codes must be borrowed. In this work we employ
results obtained in \cite{sych14}, \cite{sych15} which are close to predictions
of the \textsc{gheisha} code \cite{fes16} used in \textsc{corsika.}

The solution of the homogeneous equation 
\begin{equation}
\label{sys.8}
\frac{\partial h_{ni}(X)}{\partial X}=-h_{ni}(X)\left( \frac{1}{\lambda _{ni}}+\frac{B_{n}}{E_{i}X}\right) 
\end{equation}
 has the form
\[
h_{ni}(X_{o}+\Delta X)=h_{ni}(X_{o})\exp \left\{ \frac{-\Delta X}{\lambda _{ni}}\right\} \left( \frac{X_{o}}{X_{o}+\Delta X}\right) ^{B_{n}/E_{i}},\]
 if for simplicity we neglect the weak dependence of \( B_{n} \) on \( X \).
The solution of the full equation (\ref{sys.5}) may be written as

\begin{eqnarray}
h_{ni}(X_{o}+\Delta X) & = & h_{ni}(X_{o})\exp \left\{ -\frac{\Delta X}{\lambda _{ni}}\right\} \left( \frac{X_{o}}{X_{o}+\Delta X}\right) ^{B_{n}/E_{i}}\label{sys.9} \\
 &  & +\sum _{m}\sum _{j_{min}}^{j_{max}}\int _{X_{o}}^{X_{o}+\Delta X}h_{mj}(X')\left[ \frac{W^{ji}_{mn}}{\lambda _{mj}}+\frac{B_{m}D^{ji}_{mn}}{E_{j}X'}\right] \nonumber \\
 &  & \qquad \qquad \qquad \exp \left\{ -\frac{X_{o}+\Delta X-X'}{\lambda _{ni}}\right\} \left( \frac{X'}{X_{o}+\Delta X}\right) ^{B_{n}/E_{i}}dX'\nonumber 
\end{eqnarray}
 which may be verified directly. The above formula allows to calculate \( h_{ni} \)
at depth \( X_{o}+\Delta X \), provided that \( h_{ni}(X) \) at \( X=X_{o} \)
is known and all \( h_{mj}(X) \) are also known for \( X_{o}\leq X\leq X_{o}+\Delta X \)
and \( j>i \). So, starting from \( h_{ni}(X_{o}) \) one may sequentially
find \( h_{ni}(X_{o}+\Delta X) \) and so on. Test calculations show that for
\( E_{i}>B_{n}/3 \) it is quite sufficient to use Simpson's formula. Here,
one needs the values for \( h_{mj}(X_{o}+\Delta X/2) \), which are obtained
via interpolation. Whereas for pions this works without problem, for kaons the
requirement \( E_{i}>B_{K^{\pm }}/3 \) becomes more restrictive and, what is
especially essential, errors for one component manifest themselves in other
components. In order to retain accuracy without resorting to excessively small
\( \Delta X \), the integration over \( X' \) in eq.\ (\ref{sys.9}) is approximated
as 
\begin{equation}
\label{sys.11}
\int\limits _{X_{o}}^{X_{o}+\Delta X}f(X')\, \left( \frac{X'}{X_{o}+\Delta X}\right) ^{B_{\pi }/E_{i}}\, dX'=A_{1}\, f(X_{o})+A_{2}\, f(X_{o}+\Delta X/2)+A_{3}\, f(X_{o}+\Delta X)
\end{equation}
 and coefficients \( A_{1} \), \( A_{2} \), \( A_{3} \) are found from the
condition that (\ref{sys.11}) is exact for a second order polynomial. The accuracy
of \( \sim 1\% \) may be achieved with \( \Delta X\simeq 5 \)~g/cm\( ^{2} \).

Other EAS characteristics (e.g. electron and muon numbers) are computed in a
traditional way as corresponding functionals from functions \( h_{n}(E,X). \)
Usually one assumes that neutral pions decay immediately at the generation point
and do not contribute to the development of the hadronic cascade. This assumption
is quite adequate at energies below the corresponding decay constant which for
neutral pions is about \( 3\cdot 10^{19} \)eV. Moreover, as primary particles
are nucleons there is an additional factor of \( \sim  \)10 in our favour.
So the number of neutral pions produced at depth \( X \) may be obtained as
\begin{equation}
\label{hpi}
h_{\pi ^{0}}(E_{i},X)=\sum _{m}\sum ^{j_{max}}_{j=i}h_{mj}(X)\left[ \frac{1}{\lambda _{mj}}W^{ji}_{m\pi ^{0}}+\frac{B_{m}}{E_{j}X}D^{ji}_{m\pi ^{0}}\right] 
\end{equation}
 where we substitute \( n \) by \( \pi ^{0} \) in the second term of the right-hand
side of eq.\ (\ref{sys.5}). The electron number \( N_{e} \) at depth \( T \)
is given as

\begin{equation}
\label{ne}
N_{e}(T)=\sum ^{i_{max}}_{i=1}\int ^{T}_{0}h_{\pi ^{0}}(E_{i},X_{0})\, N_{G}(y,t)\, \frac{2}{1+s}\, dX_{0}
\end{equation}
 where \( y=\ln (E_{i}/\beta ) \) is the logarithmic energy in units of the
critical energy of electrons in air (\( \beta =81.10^{6} \) eV) and \( t=(T-X_{0})/T_{0} \)
is the depth difference in radiation units (\( T_{0}= \)37.1 g/cm\( ^{2} \)).
The factor \( 2/(1+s) \) accounts for energy sharing between two photons, with
\( s \) being the shower age parameter \( s=3t/(t+2y) \). The function \( N_{G} \)
is referred to as Greisen's formula \cite{gre17}, 

\begin{equation}
\label{ng}
N_{G}=\frac{0.31}{\sqrt{y}}\exp \left\{ t\, \left[ 1-\frac{3}{2}\ln (s)\right] \right\} ,
\end{equation}
 which predicts the electron number at depth \( T \) in the shower, produced
by a primary photon with energy \( E_{i} \) at depth \( X_{0} \).

As a rule, experimental EAS arrays can detect muons with energies above a certain
threshold \( E_{\mathrm{thr}\, \mu } \). The number of such muons may be obtained
as follows:

\begin{equation}
\label{nm}
N_{\mu }(E_{\mu }>E_{\mathrm{thr}\, \mu },T)=\sum _{m}\sum _{i:E_{i}>E_{\mathrm{thr}\, \mu }}\sum _{j\geq i}\int _{0}^{T}h_{mj}(X)\; \frac{B_{m}}{E_{j}X}\; D_{m\mu }^{ji}\; W(E_{i},X,E_{\mathrm{thr}\, \mu },T)\, dX,
\end{equation}
 where \( D_{m\mu }^{ji} \) defines the number of muons with energy \( E_{i} \)
resulting from the decay of hadron \( m \) with energy \( E_{j} \), \( W(E_{i},X,E_{\mathrm{thr}\, \mu },T) \)
is the probability that a muon produced at depth \( X \) with energy \( E_{i} \)
will survive between \( X \) and \( T \) and its final energy at \( T \)
will be greater than \( E_{\mathrm{thr}\, \mu } \).

For the most important channels of muon production (\( \pi ^{\pm }\rightarrow \mu ^{\pm }+\nu _{\mu } \)
and \( K^{\pm }\rightarrow \mu ^{\pm }+\nu _{\mu } \) are responsible for about
95\% of all muons) values of \( D_{m\mu }^{ji} \) are governed by the simple
two body decay kinematics. If we consider the atmosphere to be isotermic then
the function \( W(E_{i},X,E_{\mathrm{thr}\, \mu },T) \) may be written explicitly:
\begin{equation}
\label{w}
W(E_{i},X,E_{\mathrm{thr}\, \mu },T)=\theta \left( E_{i}-E_{\mathrm{thr}\, \mu }-(T-X)\frac{dE}{dX}\right) \left\{ \frac{X}{T}\frac{E_{i}-(T-X)\frac{dE}{dX}}{E_{i}}\right\} ^{B_{\mu }/(E_{i}+X\frac{dE}{dX})}
\end{equation}
 where \( \theta  \) is the step function and \( \frac{dE}{dX} \) is the ionization
loss rate.

\section{Calculations of the Inclusive Spectra \label{cins.label}}

In this section, we discuss how to obtain the matrices \( W_{mn}^{ji} \), representing
inclusive particle spectra, based on \textsc{neXus} calculations.

The first step evidently amounts to performing a number of \textsc{neXus} simulations
to calculate the energy spectrum of produced (secondary) hadrons of type \( h \),
\begin{equation}
E\frac{dN_{h_{\mathrm{in}}h}}{dE}(E_{\mathrm{in}},E),
\end{equation}
 in a hadron plus nitrogen reaction \( h_{\mathrm{in}}\mathrm{N}\rightarrow hX \)
for different incident hadrons \( h_{\mathrm{in}} \), each one at different
energies \( E_{\mathrm{in}} \). For the incident energies, we use
\begin{equation}
E^{j}_{\mathrm{in}}=10^{2+j/2}\mathrm{GeV},\quad j=0,1,\ldots 12,
\end{equation}
and for the incident hadron types, we take nucleons, charged pions, charged
kaons, and neutral kaons \( K^{0}_{L} \), for the secondary ones in addition
neutral pions and \( K^{0}_{S} \) (spectra of \( K^{0}_{S} \) are assumed
to be identical to \( K^{0}_{L} \)). All other hadrons are assumed to decay
immediately. 

\textsc{neXus} is based on the Monte Carlo technique, implying automatically
statistical fluctuations, which are in particular large, when the production
of secondaries with energies approaching the primary one is examined. The influence
of the limited statistics obtained via \textsc{neXus} Monte Carlo simulations
may be eliminated to some extent if an appropriate smoothing procedure is applied.
Such a procedure provides the opportunity to exploit the corresponding `smoothened'
spectrum at a fixed primary energy as a continuous function of the energy \( E \)
of secondaries, instead of considering discrete values only. Moreover, it enables
one to impose certain restrictions on the shape of inclusive spectra. For example,
we assume the \( E \) dependence of \( W \) for \( x=E/E_{\mathrm{in}}\rightarrow 1 \)
to be proportional to \( x^{-1}(1-x)^{\alpha } \), where \( a \) is taken
from theoretical considerations or just as a fit parameter. We use the Levenberg-Marquardt
(LM) method \cite{LM18} to fit the Monte Carlo spectra by analytic continuous
functions in \( E \),
\begin{equation}
E\frac{dN_{h_{m}h_{n}}}{dE}(E^{j}_{\mathrm{in}},E)\rightarrow W_{mn}(E^{j}_{\mathrm{in}},E)
\end{equation}
for the different incident hadrons \( h_{m} \) and secondaries \( h_{n} \)
for the above-mentioned values \( E^{j}_{\mathrm{in}} \) for the incident energy.
This method finds the minimum \( \chi ^{2} \) and its algorithm consists in
the combination of the inverse-Hessian method and the steepest descent method.
It is one of the standard non-linear least-square approaches and its detailed
description may be found elsewhere \cite{WHP19}. The statistics of \( 10^{5} \)
events for each of the reference energies \( E^{j}_{\mathrm{in}} \) is sufficient
to calculate \( W_{mn}(E^{j}_{\mathrm{in}},E) \) within \( \sim (0.1\div 0.2)\% \)
accuracy. 

\begin{figure}[htb]
{\par\centering \resizebox*{!}{0.9\textheight}{\includegraphics{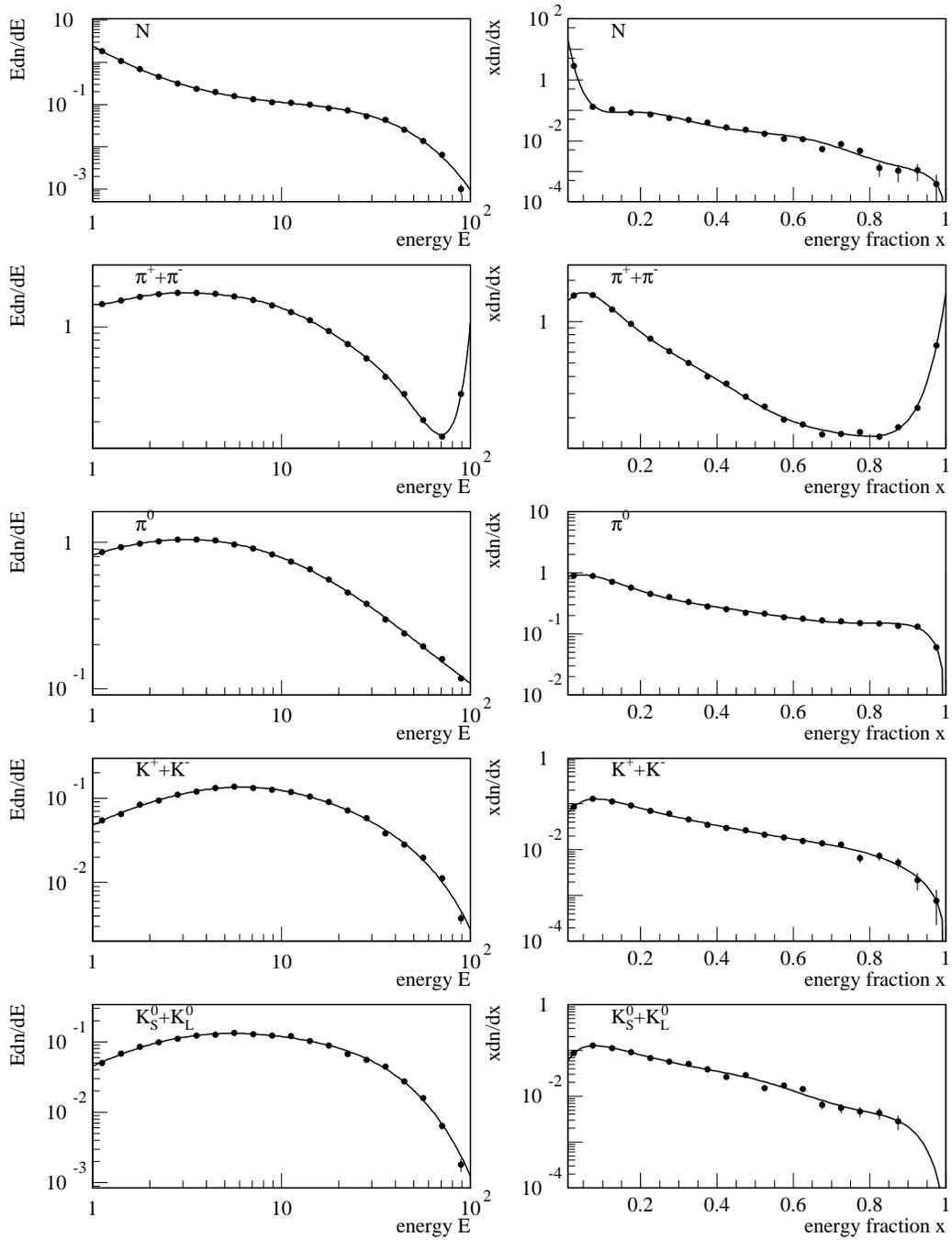}} \par}

\caption{Inclusive spectra (points) from \textsc{neXus} simulations and the corresponding
smoothened spectra \protect\( W_{mn}(E^{j}_{\mathrm{in}},E)\protect \) (lines)
for incident charged pions and different secondaries \protect\( h_{n}\protect \),
for incident energy \protect\( E_{\mathrm{in}}=10^{11}\, \protect \)eV, as
a function of the secondary energy \protect\( E\protect \) (left) and \protect\( x=E/E_{\mathrm{in}}\protect \)
(right). \label{1}}
\end{figure}

\begin{figure}[htb]
{\par\centering \resizebox*{!}{0.9\textheight}{\includegraphics{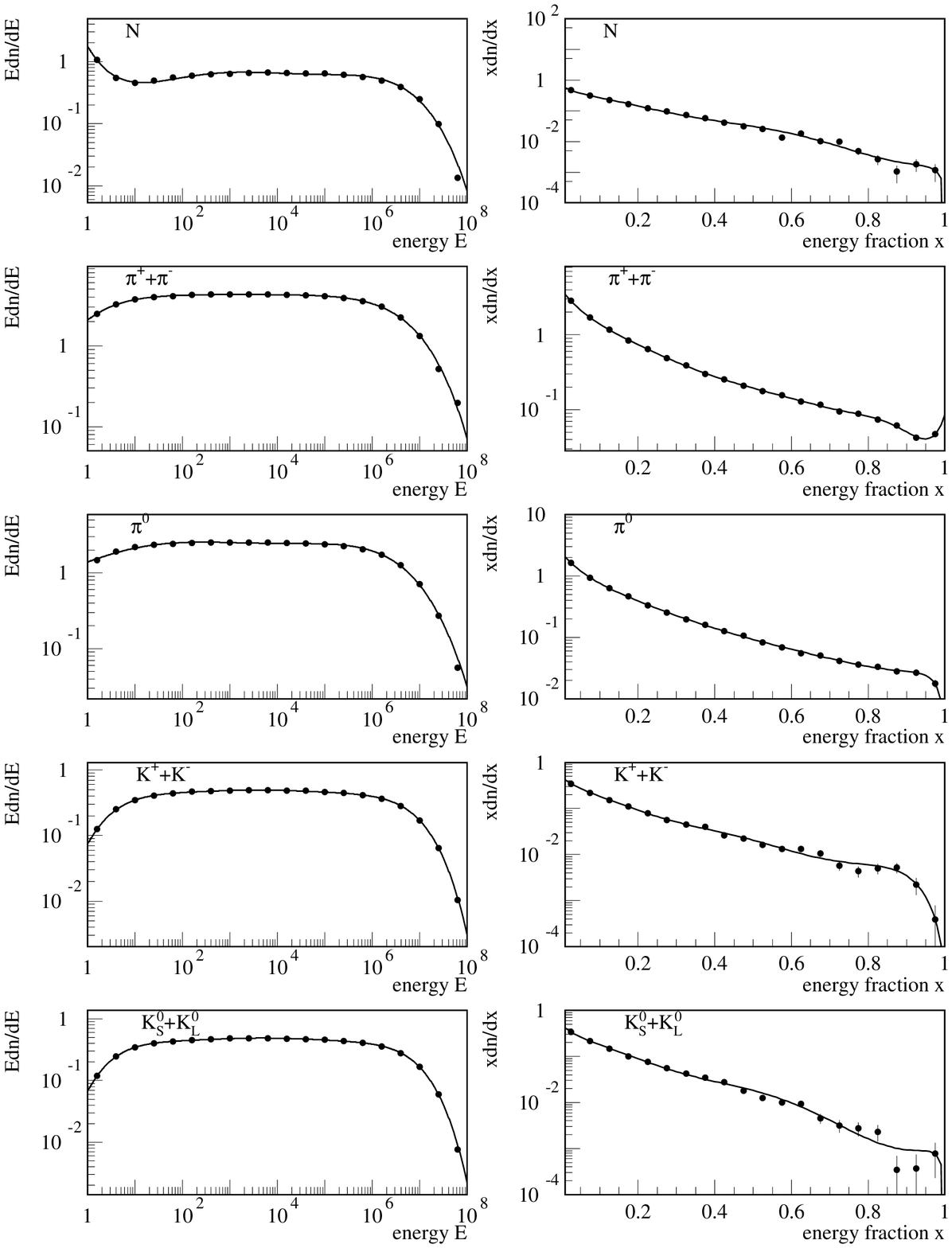}} \par}

\caption{Same as fig. \ref{1}, but for incident energy \protect\( E_{\mathrm{in}}=10^{17}\, \protect \)eV.\label{3}}
\bigskip{}
\end{figure}

As an example, we show in figs.\ \ref{1} and \ref{3}
the Monte Carlo results for incident charged pions with energies of \( 10^{11} \)
eV and \( 10^{17} \) eV, together with the corresponding fit functions \( W_{mn}(E^{j}_{\mathrm{in}},E) \).
Similarly excellent fits are obtained for all the other spectra. For purposes
of more efficient computing, two sets of Monte Carlo spectra are used, which
exploit different \( E \)-scales: a linear one for large energies of secondaries
(right figures) and a logarithmic scale for small ones (left figures).

Using analytic expressions for the spectra, we can now proceed to calculate
integrated spectra,
\begin{equation}
\label{wjmn}
W^{i}_{mn}(E^{j}_{\mathrm{in}})=\int _{E_{i}/\sqrt{c}}^{E_{i}\cdot \sqrt{c}}\frac{E}{E_{i}}W_{mn}(E^{j}_{\mathrm{in}},E)dE
\end{equation}
 with \( E_{i}=E_{\mathrm{min}}\, \mathrm{c}^{\mathrm{i}} \) for \( 1\leq i\leq i_{\mathrm{max}} \).
For the actual calculation, we use \( c=10^{0.1} \) and \( i_{\mathrm{max}}=81 \).
We obtain the integrated spectra for arbitrary incident energies \( E_{\mathrm{in}} \)
via an interpolation formula \( \tilde{W}^{i}_{mn}(E_{\mathrm{in}}) \), and
we can thus calculate these quantities in particular for all the energies \( E_{j} \):
\begin{equation}
W^{ji}_{mn}=\tilde{W}^{i}_{mn}(E_{j}).
\end{equation}
One could have calculated the integrated spectra directly for all the energies
\( E_{j} \), but due to the weak energy dependence of the spectra it is much
more efficient to proceed as discussed above.

\section{Solving the Cascade Equations}

Having all the ingredients, we are now able to solve the cascade equations as
discussed above to obtain the differential hadron spectra \( h_{n}(E_{i},X) \).
The method adopted gives the possibility to calculate average characteristics
of EAS within \( \sim 1 \)\% accuracy. Based on the inclusive spectra, we may
calculate numbers of different hadrons \( N_{h_{i}} \) as well as numbers of
electrons \( N_{e} \) and \( N_{\mu } \) for a number of observation levels
\( X \) in the atmosphere.
\begin{figure}[htb]
{\par\centering \resizebox*{!}{0.4\textheight}{\includegraphics{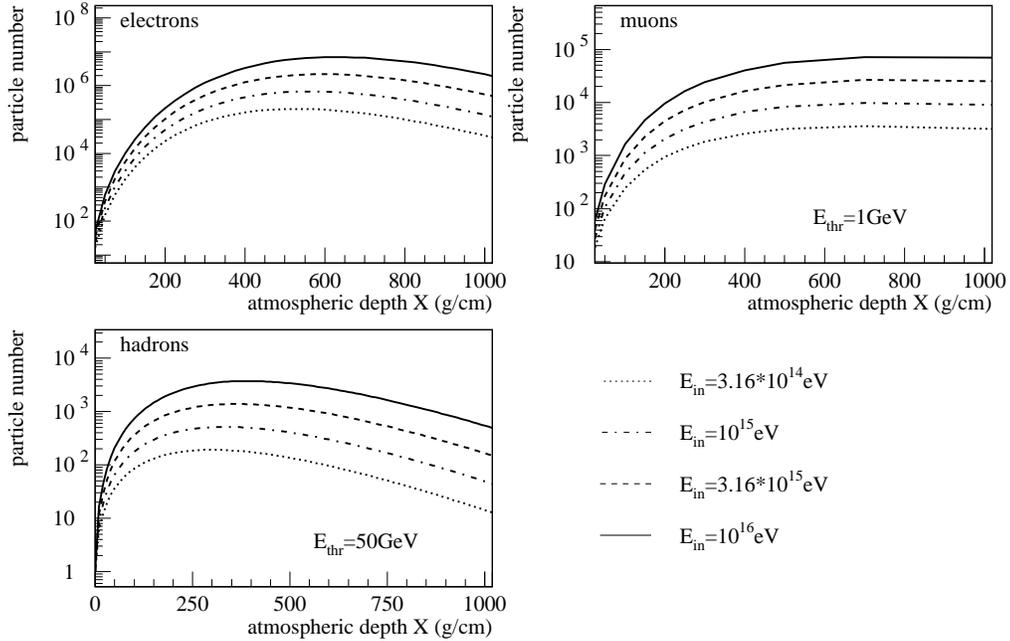}} \par}

\caption{Number of electrons \protect\( N_{e}\protect \), muons \protect\( N_{\mu }\protect \)
(\protect\( E_{\mu }>1\protect \) GeV), and all hadrons \protect\( N_{h}\protect \)
(\protect\( E_{h}>50\protect \) GeV) as a function of the atmospheric depth
\protect\( X\protect \) for different incident energies \protect\( E_{\mathrm{in}}\protect \)
(in eV, from top to bottom): \protect\( 10^{16}\protect \), \protect\( 10^{15.5}\protect \),
\protect\( 10^{15}\protect \), \protect\( 10^{14.5}\protect \).\label{rates}}
\end{figure}
\begin{figure}[htb]
{\par\centering \resizebox*{!}{0.59\textheight}{\includegraphics{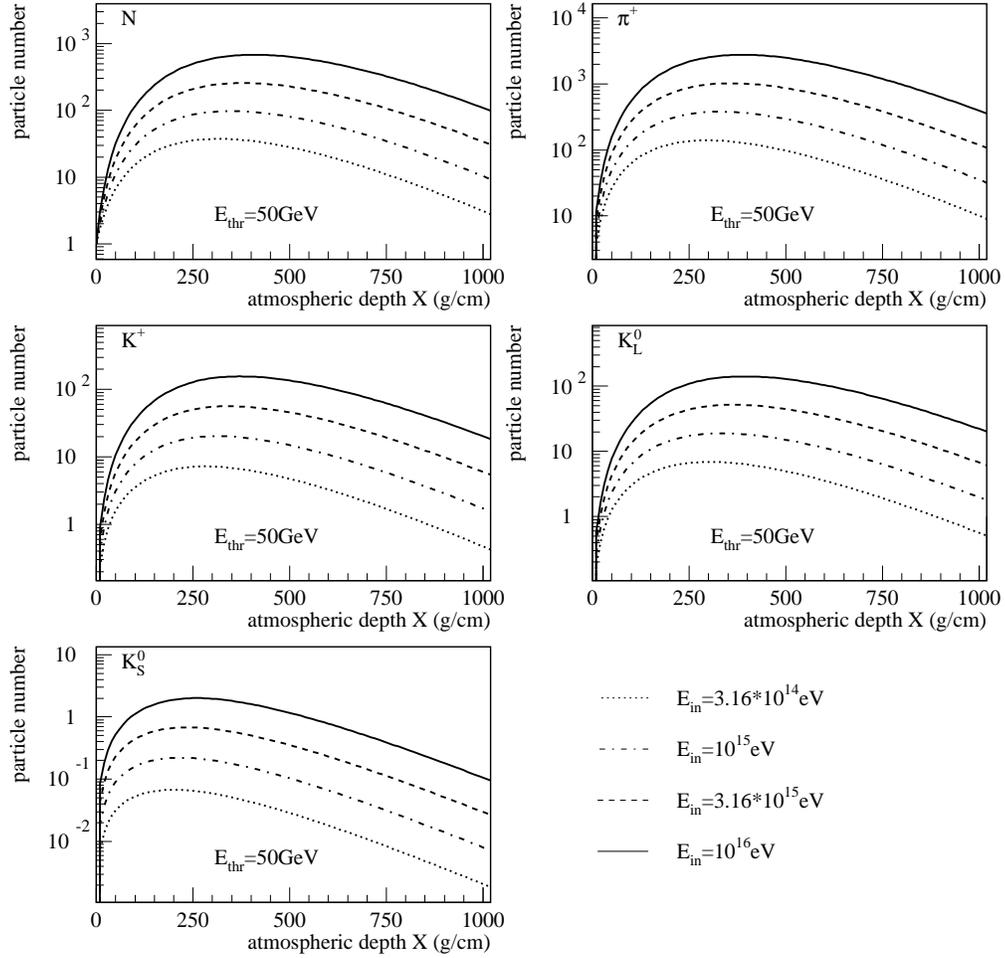}} \par}

\caption{Numbers of different hadrons \protect\( N_{h_{i}}\protect \) (\protect\( E_{h_{i}}>50\protect \)
GeV) as a function of the atmospheric depth \protect\( X\protect \) for different
incident energies \protect\( E_{\mathrm{in}}\protect \). \label{irates}}
\end{figure}
\begin{figure}[htb]
{\par\centering \resizebox*{!}{0.24\textheight}{\includegraphics{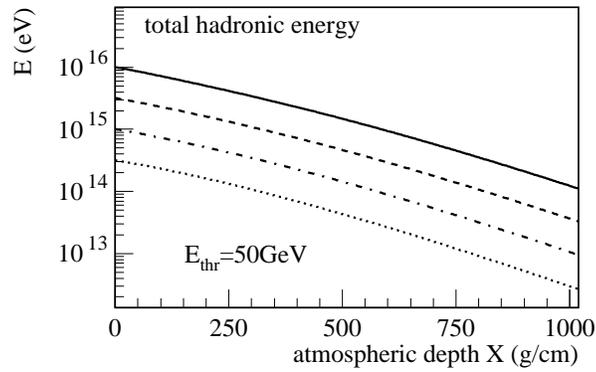}} \par}

\caption{Total hadronic energy \protect\( E\protect \) as a function of the atmospheric
depth \protect\( X\protect \) for different incident energies (in eV, from
top to bottom): \protect\( 10^{16}\protect \), \protect\( 10^{15.5}\protect \),
\protect\( 10^{15}\protect \), \protect\( 10^{14.5}\protect \).\label{energy}}
\end{figure}
In fig.\ \ref{rates}, we show the dependencies of the number of electrons \( N_{e} \),
muons \( N_{\mu } \) (\( E_{\mu }>1 \) GeV), and all hadrons \( N_{h} \)
(\( E_{h}>50 \) GeV) on the depth \( X \) for different incident energies.
One observes the expected increase of the particle numbers with energy and as
well the shift of the shower maximum towards larger \( X \) with increasing
energy. In fig.\ \ref{irates}, we show the corresponding characteristics for
individual hadrons. The pions are by far most dominant, followed by nucleons,
then charged kaons and \( K^{0}_{L} \), whereas \( K^{0}_{S} \) are the least
frequent due to the short life time. In fig.\ \ref{energy}, we show the total
hadronic energy as a function of the atmospheric depth \( X \) for different
incident energies. Obviously the hadronic energy is highest for the highest
incident energy. With increasing atmospheric depth, the hadronic energy drops
exponentially, due to its conversion into the energy of electro-magnetic cascade
(and to some extent into muon and neutrino energy).

\section{Some Results of Calculations\label{src.label}}

It is certainly the main purpose of this paper to introduce a new, sophisticated
hadronic interaction model, particularly suited for high energy hadronic interactions,
and, at the same time, to explain the use of cascade equations rather than explicit
simulations for the air shower calculation. So we do not want to present extensive
applications of this approach, but rather discuss one instructive example.

\begin{figure}[htb]
{\par\centering \resizebox*{!}{0.3\textheight}{\includegraphics{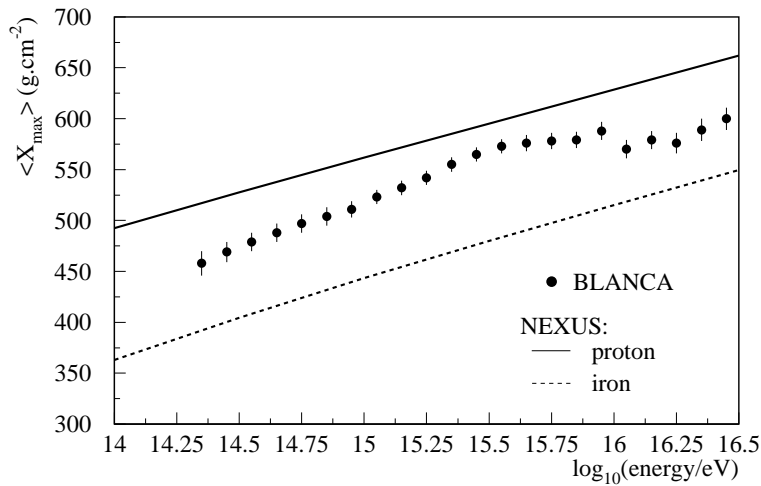}} \par}

\caption{The shower maximum \protect\( X_{\mathrm{max}}\protect \) for incident nucleons
(upper curve) and iron (lower curve) as a function of the incident energy \protect\( E_{\mathrm{in}}\protect \)
together with the data \cite{C-B20}.\label{max}}
\end{figure}

The \textsc{casa-blanca} group \cite{C-B20} published recently very interesting
results concerning the composition of cosmic rays in the energy range \( 10^{14} \)-\( 10^{17} \)
eV. From the results shown in the previous section, we can easily calculate
the shower maximum \( X_{\mathrm{max}} \) as a function of the incident energy
\( E_{\mathrm{in}} \). We calculate as well the shower maximum for incident
nuclei (with mass number \( A \)) assuming that it is given by the result for
nucleon at a reduced energy \( E_{\mathrm{in}}/A \). As it had been shown (see,
for example, \cite{KaO21,qgs02}) this simple superposition prescription works
well for average characteristics of nucleus-induced EAS. In fig.\ \ref{max}
we show the shower maximum \( X_{\mathrm{max}} \) for incident nucleons (upper
curve) and iron (lower curve) as a function of the incident energy \( E_{\mathrm{in}} \)
together with the data. Comparing our calculations with the data we find an
excellent agreement with the results of analysis carried out in \cite{kam22}
on the basis of \textsc{qgsjet} and \textsc{venus} models, i.e.\ we confirm
the change to a heavier primary composition at energies above the knee of the
primary cosmic ray spectrum, which has been earlier reported by the group of
Moscow State University \cite{khr94,fom23}, as well as the novel feature discovered
by the \textsc{casa-blanca} group - ``lightening'' of the composition at energies
just before the knee.

\begin{figure}[htb]
{\par\centering \resizebox*{!}{0.3\textheight}{\includegraphics{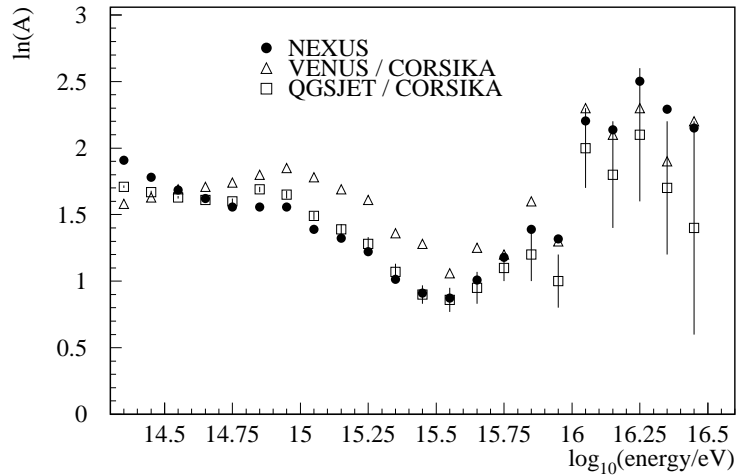}} \par}

\caption{The mean logarithmic mass ln\protect\( \left\langle A\right\rangle \protect \)
as a function of the incident energy \protect\( E_{\mathrm{in}}\protect \).
We compare our \textsc{neXus} results with the \textsc{qgsjet} and \textsc{venus}
ones obtained in \cite{C-B20}. \label{lna}}
\end{figure}

The above statement can be made more quantitative by studying the so-called
mean nuclear mass \( \left\langle A\right\rangle  \), which is defined to be
the nuclear mass which would fit best the experimental data. In fig.\ \ref{lna},
we plot the mean logarithmic mass ln\( \left\langle A\right\rangle  \) as a
function of the incident energy \( E_{\mathrm{in}} \) together with \textsc{qgsjet}
and \textsc{venus} results obtained in \cite{C-B20}. As already mentioned above,
the mass number has clearly a minimum at around \( 10^{15.5} \)\( \,  \)eV,
for higher energies the mass is increasing again. The \textsc{neXus} results
are quite similar to the \textsc{qgsjet} ones, whereas \textsc{venus} has a
tendency towards higher masses. One should keep in mind, however, that \textsc{venus}
is strictly speaking already outside its energy range of validity, which makes
its prediction somewhat uncertain.

It is worth to note that the observed qualitative behaviour for the primary
composition may be obtained in the framework of the diffusion model for cosmic
ray propagation if one assumes a large magnetic halo for the Galaxy with antisymmetric
magnetic field \cite{ptu24}. As it has been shown in \cite{KaP25} such a configuration
of the galactic magnetic field agrees with the measurements and allows to explain
naturally the observed ``sharpness'' of the knee in the primary energy spectrum.

\section{Summary and Outlook}

We introduced a new type of hadronic interaction model (\textsc{neXus}), which
has a much more solid theoretical basis when compared, for example, to presently
used models such as \textsc{qgsjet} and \textsc{venus}, and so provides much
more reliable predictions at superhigh energies where there are no collider
data yet. A particular feature of the model is a consistent treatment for calculating
cross sections and particle production considering energy conservation strictly
in both cases. In addition, one introduces hard processes avoiding any unnatural
dependence on artificial cutoff parameters. Using a single set of parameters,
\textsc{neXus} is able to fit many basic spectra in proton-proton and lepton-nucleon
scattering, as well as in electron-positron annihilation. So it is worth to
point out once more that concerning theoretical consistency, \textsc{neXus}
is considerably superior to the models in current use, and allows a much safer
extrapolation to superhigh energies which are very important in cosmic ray studies.

We explained in detail an air shower calculation scheme, based on cascade equations,
which is quite accurate in calculating main characteristics of air showers,
being extremely fast concerning computing time. We perform the calculation using
the \textsc{neXus} model to provide the necessary tables concerning particle
production in hadron-air collisions.

As an application, we calculated the shower maximum as a function of the incident
energy for incident protons and iron, to compare with corresponding data. Based
on these data, we calculated as well the so-called mean nuclear mass. We are
thus able to confirm that the average mass as a function of the incident energy
shows a minimum around \( 10^{15.5} \)\( \,  \)eV.

Problems where it is possible to ignore fluctuations are not too numerous. So
an unavoidable question arises how to retain calculation efficiency at reasonable
level and at the same time to account for fluctuations. But the answer to this
really crucial question is well known and proves to be rather simple. One needs
to employ a combination of the Monte Carlo techhnique and numerical solutions
of hadronic cascade equations in the atmosphere. The explicit simulation should
be carried out from the primary (initial) energy \( E_{\mathrm{in}} \) to some
threshold \( E_{\mathrm{thr}}=kE_{0} \), where \( k\sim 10^{-2}-10^{-3} \).
Numerous calculations showed (see, for example, \cite{qgs02}) that there is
no sense in going below this threshold as it does not increase the accuracy
of EAS fluctuation determination. So contributions from hadrons with energies
below \( E_{\mathrm{thr}} \) may be accounted for in average. Numerical solutions
of hadronic cascade equations can produce corresponding tables with sufficient
accuracy and in a very short time.

Here we do not consider calculations of lateral distributions of shower particles
as this problem will be discussed in our next paper. But we would like to point
out that there is a rather important class of problems connected with giant
EAS arrays aimed at primary energies \( \geq 10^{20} \)eV (see \cite{AUG26}).
For these arrays it is sufficient to calculate lateral distributions only at
large distances (above 100 m) from the shower axis. Such a situation makes it
possible to treat the problem as a combination of the one-dimensional approach
for hadrons with energy \( \geq 10^{11} \)eV and the rigorous three-dimensional
technique for low energy region only.

\section{Acknowledgements}

This work has been funded in part by the IN2P3/CNRS (PICS 580) and the Russian
Foundation of Basic Researches (RFBR-98-02-22024). 

\bibliographystyle{pr2}
\bibliography{cosmic}

\end{document}